\documentclass[letterpaper,english,reprint, aps]{revtex4-1}
\usepackage[T1]{fontenc}
\usepackage[latin9]{inputenc}
\usepackage{textcomp}
\usepackage{amsmath}
\usepackage{amssymb}
\usepackage{graphicx}
\PassOptionsToPackage{version=3}{mhchem}
\usepackage{mhchem}

\let\oldAA\AA
\renewcommand{\AA}{\text{\normalfont\oldAA}}

\makeatletter

\providecommand{\tabularnewline}{\\}

\usepackage{physics}

\makeatother

\usepackage{babel}
\begin{document}
\preprint{APS/123-QED}
\title{On the Origin of Supertetragonality in $\ce{BaTiO_{3} }$ }
\author{S. Mellaerts\textsuperscript{1}, J.W. Seo\textsuperscript{2}, V.
Afanas'ev\textsuperscript{1}, M. Houssa\textsuperscript{1,3} and
J.-P. Locquet\textsuperscript{1}}
\affiliation{\textsuperscript{1}Department of Physics and Astronomy, KU Leuven,
Celestijnenlaan 200D, 3001 Leuven, Belgium, \textsuperscript{2}Department
of Materials Engineering, KU Leuven, Kasteelpark Arenberg 44, 3001
Leuven, Belgium, \textsuperscript{3}Imec, Kapeldreef 75, 3001 Leuven,
Belgium}
\email{simon.mellaerts@kuleuven.be}

\begin{abstract}
Understanding ferroelectricity is of both fundamental and technological
importance to further stimulate the development of new materials designs
and manipulations. Here, we perform an in-depth first-principle study
on the well-known ferroelectric barium titanate $\ce{BaTiO_{3} }$
under a hydrostatic negative pressure, showing an isosymmetric phase
transition to a supertetragonal phase with high $c/a$ ratio of $\sim1.3$.
The microscopic origin and driving mechanisms of this phase transition
are identified as a drastic change of the covalently $\pi$-bonded
electrons. These findings provide guidance in the search for new supertetragonal
phases, with great opportunities for novel multiferroic materials;
and can be generalized in the understanding of other isosymmetric
phase transitions.
\end{abstract}
\maketitle

\section{\label{sec:Intro}Introduction}

Complex metal oxides host a rich variety of functional properties
- including multiferroicity, dielectricity, metal-insulator transitions
and topological phases - that can be controlled by doping and epitaxial
strain \citep{JP_doublingTc_LSrCuO,Pia_pinningMott}. This tunability
has strongly motivated the technological interest in these metal oxides,
and although past decades have witnessed great efforts in fundamental
research, there remains a great need for novel materials designs and
manipulations. The former has been an essential pathway to find new
multiferroic systems, and to the exploration of double perovskites
\citep{doubleperovs_Dasgupta,doubleperovs_Cai} as well as the idea
of ferroelectric materials driven by polar octahedral rotations \citep{BENEDEK201211,Fennie_octahedralrotations},
while for the latter the development of vertically aligned nanocomposites
(VANs) has recently attracted great attention \citep{VAN_first,VAN_summaryDriscoll,VAN_superconductivity,VAN_ETO,Zhang_NPenhancedTc_BTO,Driscollreview}.
For example, an enhanced Curie temperature of $616^\circ$C in $\ce{BaTiO_{3} }$
has been induced through negative pressure \citep{Zhang_NPenhancedTc_BTO}.

This experimental effort towards three-dimensional (3D) strain opens
a largely unexplored region in phase space, experimentally as well
as computationally. Ultimately, it might open the path in fulfilling
the dream of having a complete atomic control of the transition metal
within the oxygen octahedron and its corresponding distortions.

\ 

In this work, we study $\ce{BaTiO_{3} }$ as a model systemand perform a
first-principle study under hydrostatic negative pressure. $\ce{BaTiO_{3} }$
is particularly interesting owing to its excellent lead-free ferroelectricity
as well as piezoelectricity. Its possibility to grow epitaxially on silicon substrates
using $\ce{SrTiO_{3} }$ buffer layers \citep{McKee1998,Norga2006,Marchiori2006} enables the integration into silicon platforms
which has gained significant interest \citep{Abel:2013,Abel:2019,Chen:2022}.
Our results demonstrate a first-order transition to a supertetragonal
(ST) phase with $c/a\sim1.3$, which has also been observed in $\ce{PbTiO_{3} }$
nanowires upon negative pressure \citep{Wang_NPsupertetra_PbTO,NPsupertetra_nanosizedPbTO},
as well as in highly strained $\ce{BiFeO_{3} }$ \citep{Bea_STBFO,Zhang_STBFO}.
However, a firm understanding of the chemical origin of this ST phase
has remained absent. By a detailed first-principle based analysis
on the intimate mixed ionic-covalent nature of $\ce{BaTiO_{3} }$,
we identify the key changes in the electronic structure that drive
the supertetragonal phase transition.

\section{\label{sec:groundstate}$\ce{BaTiO_{3} }$ Phases}

As a first part of this study, all four $\ce{BaTiO_{3} }$ phases
- cubic ($Pm\bar{3}m$), tetragonal ($P4mm$), orthorhombic ($Amm2$)
and rhombohedral ($R3m$) - are being studied on a structural and
dielectric level. All calculations were performed by the use of the
PBEsol functional as it has shown to be one of the most accurate exchange-correlations
functionals for the various ferroelectric perovskites \citep{ExchangeComparison_Yuk,Spaldin_STOstudy}.
The resulting properties of the high-temperature ($Pm\bar{3}m$ and
$P4mm$) and low-temperature phases ($Amm2$ and $R3m$) have been
summarized in Table \ref{table:highTphases} and \ref{table:lowTphases},
respectively.

\begin{table}[h]
\caption{A structural and dielelectric comparison of the cubic ($Pm\bar{3}m$)
and tetragonal ($P4mm$) $\ce{BaTiO_{3} }$ phases, with a comparison
to their experimental values. The optical dielectric function $\epsilon^{\infty}$,
Born effective charges of titanium $Z_{\ce{Ti }}^{*}$ and total polarization
$P_{s}$ are given.}
\label{table:highTphases}

\begin{tabular}{|c||c|c|}
\hline 
 & \multicolumn{2}{c|}{Cubic ($Pm\bar{3}m$)}\tabularnewline
\hline 
 & Theory & Experiment\tabularnewline
\hline 
\hline 
$a$ ($\AA$) & $3.985$ & $3.996$ \citep{cubic_lattice_ref}\tabularnewline
\hline 
$\epsilon_{ii}^{\infty}$ & $6.79$ & $5.40$ \citep{dielectric_cubic}\tabularnewline
\hline 
$(Z_{\ce{Ti }}^{*})_{ii}$ & $7.41$ & /\tabularnewline
\hline 
\hline 
 & \multicolumn{2}{c|}{Tetragonal ($P4mm$)}\tabularnewline
\hline 
 & Theory & Experiment\tabularnewline
\hline 
$a$ ($\AA$) & $3.968$ & $3.995$ \citep{tetra_lattice_ref}\tabularnewline
\hline 
$c$ ($\AA$) & $4.072$ & $4.034$ \citep{tetra_lattice_ref}\tabularnewline
\hline 
$(\epsilon_{11}^{\infty},\epsilon_{33}^{\infty})$ & $(6.45,5.66)$ & $(5.93,5.60)$ \citep{dielectric_tetra}\tabularnewline
\hline 
$(Z_{\ce{Ti }}^{*})_{ii}$ & $(7.05,7.05,5.61)$ & /\tabularnewline
\hline 
$P_{s}$ ($\mu$C/cm$^{2}$) & $35.7$ & $26.0$ \citep{tetra_polariz_ref}\tabularnewline
\hline 
\end{tabular}
\end{table}

For the structural parameters, an excellent agreement with the experimental
results could be achieved, while for the dielectric properties, there
exists a small overestimation attributed to the usual delocalized
tendency of the DFT exchange-correlation functionals.

\begin{table}[h]
\caption{A structural and dielelectric comparison of the low-temperature orthorhombic
($Amm2$) and rhombohedral ($R3m$) $\ce{BaTiO_{3} }$ phases, with
a comparison to their experimental values.}
\label{table:lowTphases}

\begin{tabular}{|c||c|c|}
\hline 
 & \multicolumn{2}{c|}{Orthorhombic ($Amm2$)}\tabularnewline
\hline 
 & Theory & Experiment\tabularnewline
\hline 
\hline 
$a$ ($\AA$) & $3.961$ & $3.983$ \citep{rhombo_lattice_ref}\tabularnewline
\hline 
$b$ ($\AA$) & $5.686$ & $5.674$ \citep{rhombo_lattice_ref}\tabularnewline
\hline 
$c$ ($\AA$) & $5.709$ & $5.692$ \citep{rhombo_lattice_ref}\tabularnewline
\hline 
$\epsilon_{ii}^{\infty}$ & $(6.35,6.06,5.63)$ & /\tabularnewline
\hline 
$(Z_{\ce{Ti }}^{*})_{ii}$ & $(6.94,6.42,5.58)$ & /\tabularnewline
\hline 
$P_{s}$ ($\mu$C/cm$^{2}$) & $44.8$ & /\tabularnewline
\hline 
\hline 
 & \multicolumn{2}{c|}{Rhombohedral ($R3m$)}\tabularnewline
\hline 
 & Theory & Experiment\tabularnewline
\hline 
$a$ ($\AA$) & $4.007$ & $4.004$ \citep{rhombo_lattice_ref}\tabularnewline
\hline 
$\alpha$ $(^{\circ})$ & $89.85$ & $89.84$ \citep{rhombo_lattice_ref}\tabularnewline
\hline 
$\epsilon_{ii}^{\infty}$ & $(5.95,5.95,5.95)$ & /\tabularnewline
\hline 
$\epsilon_{ij}^{\infty}$ $(i\neq j)$ & $-0.16$ & /\tabularnewline
\hline 
$(Z_{\ce{Ti }}^{*})_{ii}$ & $(6.22,6.22,6.22)$ & /\tabularnewline
\hline 
$(Z_{\ce{Ti }}^{*})_{ij}$ $(i\neq j)$ & $-0.32$ & /\tabularnewline
\hline 
$P_{s}$ ($\mu$C/cm$^{2}$) & $47.9$ & $34$ \citep{rhombo_polariz_ref}\tabularnewline
\hline 
\end{tabular}
\end{table}

The rich phase diagram of $\ce{BaTiO_{3} }$ is in stark contrast
to $\ce{PbTiO_{3} }$ that has only the tetragonal phase as ground state.
This is often ascribed to the lone pair of $6s$ electrons of $\ce{Pb }^{2+}$
overlapping with the oxygen orbitals to attain a lobe shape \citep{Atanasov:2001vo},
breaking the spherical symmetry and imposing restrictions on the $\ce{Ti }$
off-centring to one single direction. While for $\ce{BaTiO_{3} }$,
electronegativity analysis shows a $82\%$ ionic component in
the $\ce{Ba-O }$ bond \citep{Thomann_covalency}, which allows to
consider it fully ionic within a first approximation. Therefore, $\ce{Ba }^{2+}$
imposes no symmetry restriction on the $\ce{Ti }$ off-centring, giving
rise to its rich phase diagram, where the sequence of phase transitions
as a function of temperature can be conceived as a tendency to maximize
the covalent bonding of the titanium with the oxygens corners of the
octahedron.

\ 

\begin{figure}[h]
\includegraphics[scale=0.65]{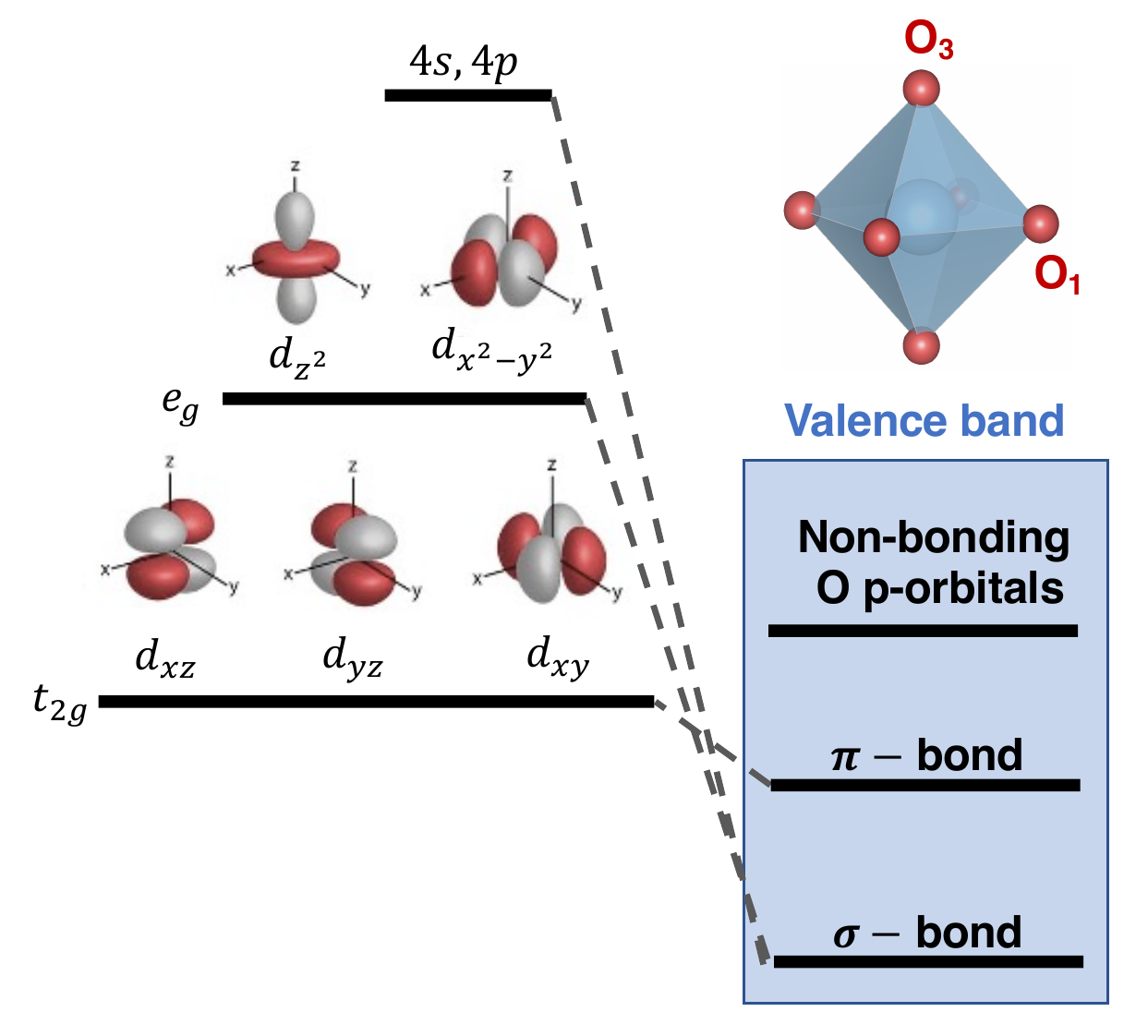}\caption{The chemical bonding in $\ce{BaTiO_{3} }$. The octahedral crystal
field splits the $3d$ orbitals into $t_{2g}$ and $e_{g}$ which
are involved in the $\sigma$- and $\pi$-bond, such that the valence
band mainly consists of the covalent bonded $\ce{Ti }$ $3d$-orbitals
and $\ce{O }$ $2p$-orbitals.}
\label{fig:chemicalbond}
\end{figure}

\begin{figure*}
\includegraphics[scale=0.63]{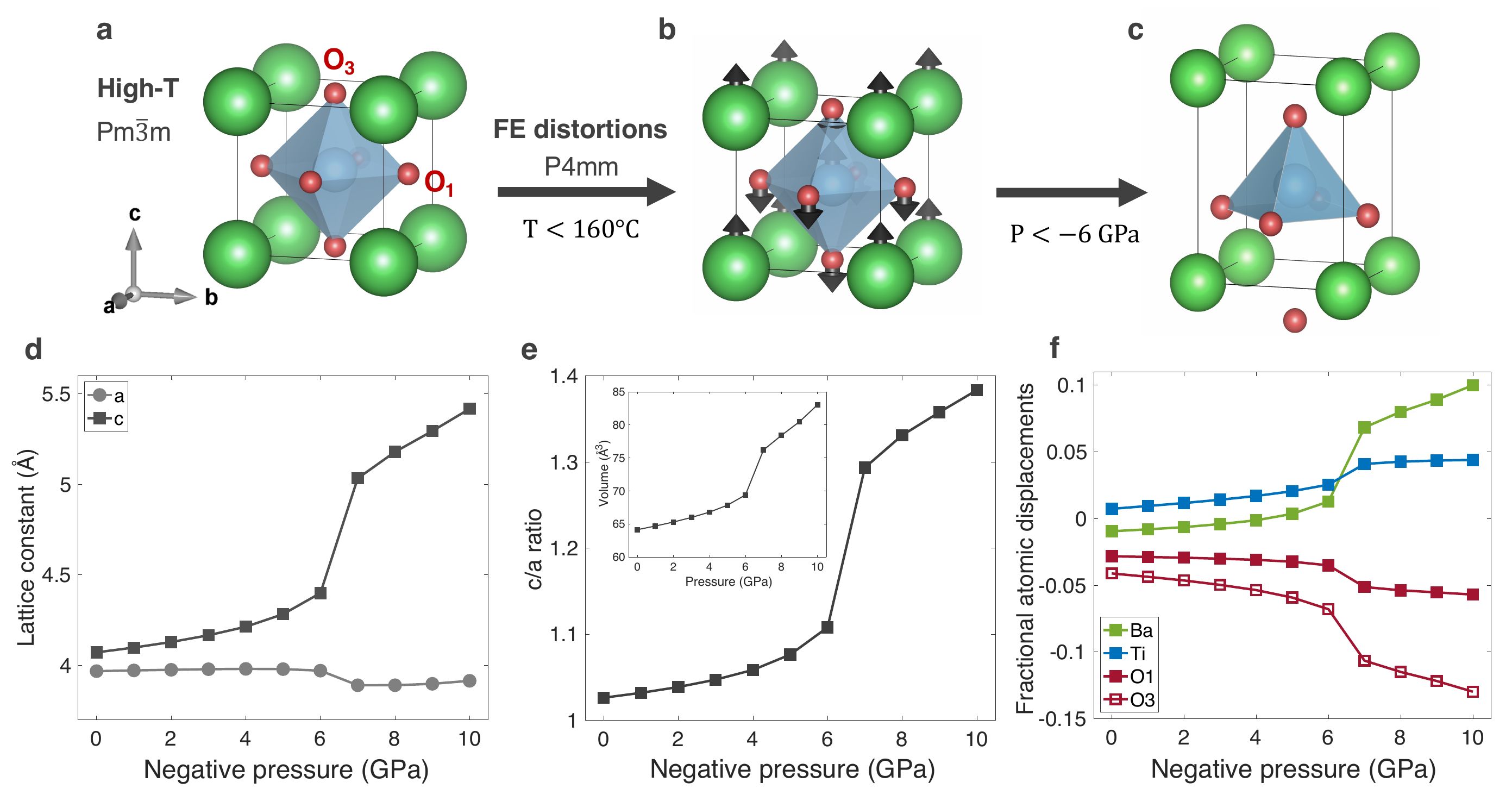}\caption{The structural component of the supertetragonal phase transition.
The crystal structures of (a) the high-temperature cubic crystal structure,
(b) the corresponding ferroelectric (FE) polar distortion, and (c)
the supertretragonal phase found in the negative pressure regime.
(d) The lattice constants show a discontinuity above $6$ GPa, which
translates into a discontinuity in (e) the $c/a$ ratio as well as
the volume (inset). (f) These discontinuities are also shown in the
fractional atomic coordinates within the unit cell.}
\label{fig:structure}
\end{figure*}

In the cubic structure, the $\ce{Ti }^{4+}$ in the oxygen octahedra
has $O_{h}$ symmetry where the crystal field splits the five-fold
degenerate $d$-orbital into a triple degenerate $t_{2g}\:(d_{xy},d_{xz},d_{yz})$
lower level and the double degenerate $e_{g}\:(d_{x^{2}-y^{2}},d_{z^{2}})$
level (see Figure \ref{fig:chemicalbond}). This latter $e_{g}$ orbital
level forms a $\sigma$-bond with all six oxygen $p$-orbitals. More
specifically, within valence band theory, these hybridize with the
$\ce{Ti }$ $4s$ and $4p$ orbital to form $sp^{3}d^{2}$ hybrid
orbital with six lobes oriented towards the linear $sp$ oxygen hybrid
orbitals, giving rise to six equivalent $\sigma$-bonds. In addition,
$\pi$ molecular orbitals are formed perpendicular to the $\ce{O-Ti-O }$
direction by the $t_{2g}$ states, which are degenerate and mutually
orthogonal. Because of the $37\%$ covalent component in the $\ce{Ti-O }$
bond \citep{Thomann_covalency}, the triple degenerate $(d_{xy},d_{xz},d_{yz})$
orbitals and their bonding dominate the ferroelectric distortion,
as they allow to weaken and overcome the spherical symmetric short-range
ionic repulsion \citep{Cohen1992}. In the cubic structure, these
orbitals form a nearly spherical symmetric delocalized electron cloud
within the octahedron, with weak negligible covalent bonds. Upon tetragonal
distortion (see Fig. \ref{fig:structure}b), the $\pi$-bond along
the polar $z$-axis will be strengthened with an increasing orbital
overlap of the $(d_{xz},d_{yz})$ orbitals with the $\ce{O_{3} }$
$(p_{x},p_{y}$) orbitals. Further lowering the temperature, forces
the $\ce{Ti }$ off-centring to the $(110)$ direction with a further
covalent enhancement of the $\pi$-bond with two oxygens in the orthorhombic
phase, and finally a $\ce{Ti }$ off-centring to the $(111)$ direction
in the rhombohedral phase with three strong $\pi$-bonds formed \citep{Smiech2000,Smiech2003,Smiech2005}.

Understanding the sequence of phase transitions within this molecular
orbital description is also informative in understanding the dielectric
trend for the different $\ce{BaTiO_{3} }$ phases. The increasing
covalent bonding tendency results in reduced dielectric constant and
Born charges upon transitioning to the low-T phases, as the charge
is no longer delocalized within the octahedron but rather forms well-defined
bonding orbitals. In other words, the transition to the lower symmetry
phase implies an enhanced covalency with well-defined directionality
- because a purely ionic compound will have the highest structural
symmetry - and with the counter-intuitive localization of the electrons
within the octahedron as the symmetry is lowered.

\section{\label{sec:Supertetragonal}Supertetragonal Phase Under Negative
Pressure}

\subsection{Structure}

A fully unconstrained atomic relaxation is performed for all four
$\ce{BaTiO_{3} }$ phases where the negative pressure has been artifically
introduced into the Cauchy stress tensor for the minimization of the
Hellmann-Feynmann forces. This implies that all structural degrees
of freedom are optimized, without imposing any crystal structure,
which is in stark contrast to a recent first-principle study \citep{PressurePerovsDFT_2020},
and more similar to the work of Tinte et al. on $\ce{PbTiO_{3} }$ \citep{VDB_supertetra}.
It can be found that upon a negative pressure above $6$ GPa, the
tetragonal phase undergoes an abrupt structural phase transition as
observed in the lattice constants and the respective $c/a$ ratio
and volume (see Figure \ref{fig:structure}d-e). The abrupt increase
has been driven by changes in the bond length, quantified by the deviations
from the high-symmetric Wyckoff positions (see Figure \ref{fig:structure}c).
Note that this phase transition does not involve any change in the
crystal symmetry, and shall therefore be referred to as isosymmetric,
which also implies a first-order nature of the phase transition \citep{Christy:ab0333}.

\ 

In addition to the ionic relaxation of the tetragonal phase upon negative
pressure, the cubic, orthorhombic and rhombohedral phases were also
structurally optimized for each pressure value (see Supplemental Information).
However, none of these phases showed any structural anomaly and therefore
were not studied further. Nonetheless, by a comparison of the enthalpy
of the different $\ce{BaTiO_{3} }$ phases, it becomes clear that
the ST phase will have the lowest enthalpy compared to the others,
due to the anomalous volume enhancement. Note that the choice of enthalpy
in the energetic comparison is considered approriate in the study
of pressure-induced phase transitions, as also shown earlier \citep{CeriumVolumeCollapse}.

\begin{figure*}
\includegraphics[scale=0.63]{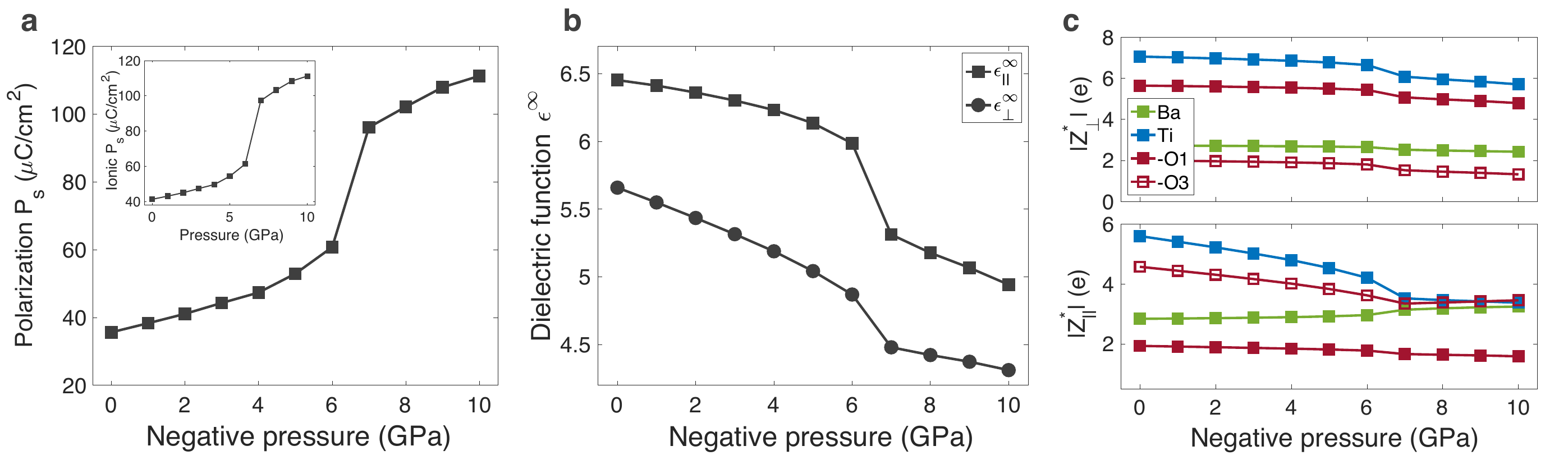}\caption{The dielectric component of the supertetragonal phase transition.
(a) The polarization as a function of pressure whose abrupt enhancement
fully arises from the ionic component (inset). (b) The dielectric
function, decomposed into a longitudinal ($\epsilon_{||}^{\infty}$)
and transversal in-plane ($\epsilon_{\perp}^{\infty}$) diagonal component.
(c) The effective Born charges of each element with its transverse
($Z_{\perp}^{*}$) and longitudinal ($Z_{||}^{*}$) diagonal components
shown.}
\label{fig:dielectric}
\end{figure*}

\subsection{\label{subsec:Dielectric-Properties}Dielectric Properties}

For each optimized pressure state, the spontaneous polarization ($P_{s}$)
was calculated within the Berry phase theory of polarization \citep{Polarization1,Polarization2},
as shown in Figure \ref{fig:dielectric}a. The anomalous enhancement
is again observed for $P_{s}$, moreover, it fully arises from the
ionic dipole moment that has been induced by the structural anomaly.
To obtain further insight into the dielectric behavior of the ST transition,
the optical dielectric tensor ($\epsilon^{\infty}$) and Born effective
charges ($Z^{*}$) were evaluated.

The transition to a ST phase results in the reduction of the dielectric
function, with a manifest discontinuity in the longitudinal component
$\epsilon_{||}^{\infty}$. The reduction is more pronounced in comparison
to the transition to the low-T orthorhombic and rhombohedral phases.
Similarly, for the Born charges an abrupt reduction is observed at
the transition. More specifically, it can be seen that $Z_{||}^{*}$
of titanium drastically reduces from the anomalous value of $5.61$
to below its atomic charge of $4$. Both trends suggest an enhanced
covalency with electrons forming well-defined localized hybrid orbitals
with less electron flow and thus a reduced polarizability, as discussed
in Sec. \ref{sec:groundstate}. The negative pressure induced electron
localization within strong covalent bonds can also be understood in
a simplistic picture where the higher volume dependence of the kinetic
energy ($V^{-2/3}$) in comparison to the electrostatic potential
energy ($V^{-1/3}$) plays a crucial role in the ST transition that
involves a large volume enhancement \citep{chemistry_under_pressure}.

\subsection{Comparison to Biaxial Strain}

\begin{figure}[h]
\includegraphics[scale=0.35]{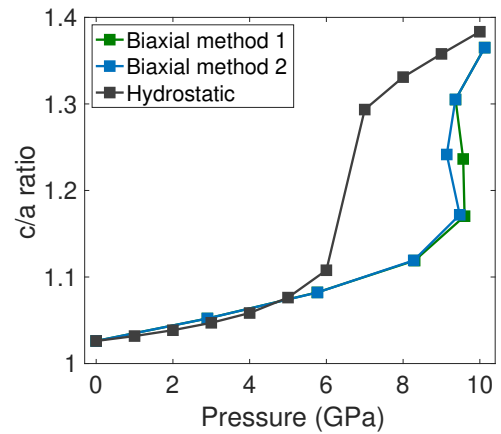} \caption{Comparison between hydrostatic and in-plane biaxial strain. The first
method (green) corresponds to an ionic relaxation with in-plane atomic
coordinates and lattice constants fixed, while the second method optimizes
all three spatial coordinates of the atoms, while fixing the in-plane
lattice constants.}
\label{fig:biaxial}
\end{figure}

In addition to the appearance of a ST phase transition upon hydrostatic
negative pressure, it was verified whether an in-plane biaxial strain
could result in a similar phase transition. To apply a biaxial strain,
two different methods were adopted; in both cases, the in-plane lattice
component were fixed, while the out-of-plane component is fully relaxed.
In the first method (1), the $(x,y)$ coordinates of the atoms have
been fixed, whereas in the second method (2) - considered as the best
approximation to the experimental reality - the atoms have been fully
relaxed in all three spatial coordinates. The comparison between both
biaxial strain methods and the hydrostatic pressure is shown in Figure
\ref{fig:biaxial}, where for the biaxial strain, the pressure is
corresponding to the in-plane stress that has been induced by the
applied compressive strain ranging from $0$ to $-7\%$.

\ 

First of all, the induced in-plane stress for the biaxial method $2$
is slightly lower in the transition with a larger $c$, which indicates
that the atoms attempt to accomodate the in-plane stress by a reordering
in which there is an increased out-of-plane distortion. 

Secondly, it can be seen that the hydrostatic pressure is much more
effective to induce the ST phase as it requires $\sim2$ GPa lower
pressure values to be applied. However, it should be emphasized that
although both ST phases in the hydrostatic and biaxial case have a
similar $c/a$ ratio, they are very distinct in any other aspect.
The ST phase induced by a compressive biaxial strain of $-6\%$ ($a=b=3.73\,\AA$)
has $c/a=1.31$ while for the negative pressure induced ST phase with
$c/a=1.33$ at $8$ GPa has a much larger in-plane lattice constant
of $a=b=3.89\,\AA$. This also means that a smaller change in the
$c$ lattice constant has been induced by biaxial strain, resulting
in a lower polarization of $P_{s}=91.1\,\mu$C/cm$^{2}$ for $c/a=1.31$
(compared to $102.2\:\mu$C/cm$^{2}$ with $c/a=1.33$ at negative
pressure $8$ GPa). This confirms that the ST can not be fully established
by a biaxial strain, and requires an additional out-of-plane tensile
stress. In other words, the ST phase (observed upon negative pressure)
is not lying on the Poisson line with $\sigma_{z}=0$ within 6D strain
space, although there still exists a first-order structural anomaly
along this line. Moreover, it explains why the ST phase can be established
for highly biaxial strained $\ce{BiFeO_{3} }$ where the $6s$ lone
pair provides a stronger internal out-of-plane stress driving it towards
a ST structure \citep{Bea_STBFO,Zhang_STBFO}.

\section{\label{sec:Microscopic}Microscopic Origin of Supertetragonality}

The origin of these supertetragonal phases remains poorly understood,
with many authors suggesting the breaking of the bond along the tetragonal
axis \citep{VDB_supertetra,Spaldin_STBFO,Cohen_supertetra_review},
albeit ill-defined what is meant. In case of a true broken bond along
the $c$-axis of the long $\ce{Ti-O }$ bond, the coordination number
of $\ce{Ti }$ would reduce to $5$, giving rise to a layered structure.
Such square pyramidal coordinations has been found in some layered
titanium complexes with a short $\ce{Ti-O }$ bond length of $1.63-1.66\,\AA$,
which is considered as symptomatic for a double bond \citep{Roberts_fivecoordinate,recent_fivecoordinate}.
However, this is in contrast to the minimum $\ce{Ti-O }$ bond length
of $1.77\,\AA$ found in the ST phase, therefore, the proposal
of a true five-fold square pyramidal $\ce{Ti }$ coordination in the
ST phase can be disregarded, and thus the long $\ce{Ti-O }$ bond
along the polar axis can not be considered as fully broken.

\ 

Similarly, it could be argued that this change in coordination is
indicating a change in the oxidation state of $\ce{Ti }^{4+}$ to
$\ce{Ti }^{3+}$, as in $\ce{LaTiO_{3} }$. However, this would mean
that $\ce{Ti }$ has an electron configuration $3d^{1}$, with the
electron localized on $\ce{Ti }$, similar to ST $\ce{PbVO_{3} }$
\citep{Goodenough_PbMO3review}. However, such a valence transition
would result in a drastic change in de partial density of states (DOS)
of the $\ce{Ti }$ $3d$ orbitals, which is not the case. Furthermore,
by Bader charge \citep{Bader:1991} and M\"ulliken orbital population
analysis \citep{mulliken_orbitalpopulation} (see Supplemental Information),
no change in the valence state could be inferred. On the other hand,
these charge analysis are inconclusive on whether the ST transition
can be explained as a drastic change from ionic to covalent bonding,
nor does the electron localization function (ELF) shows any significant
changes upon the transition (see Supplemental Information).

\begin{figure*}[t]
\includegraphics[scale=1.10]{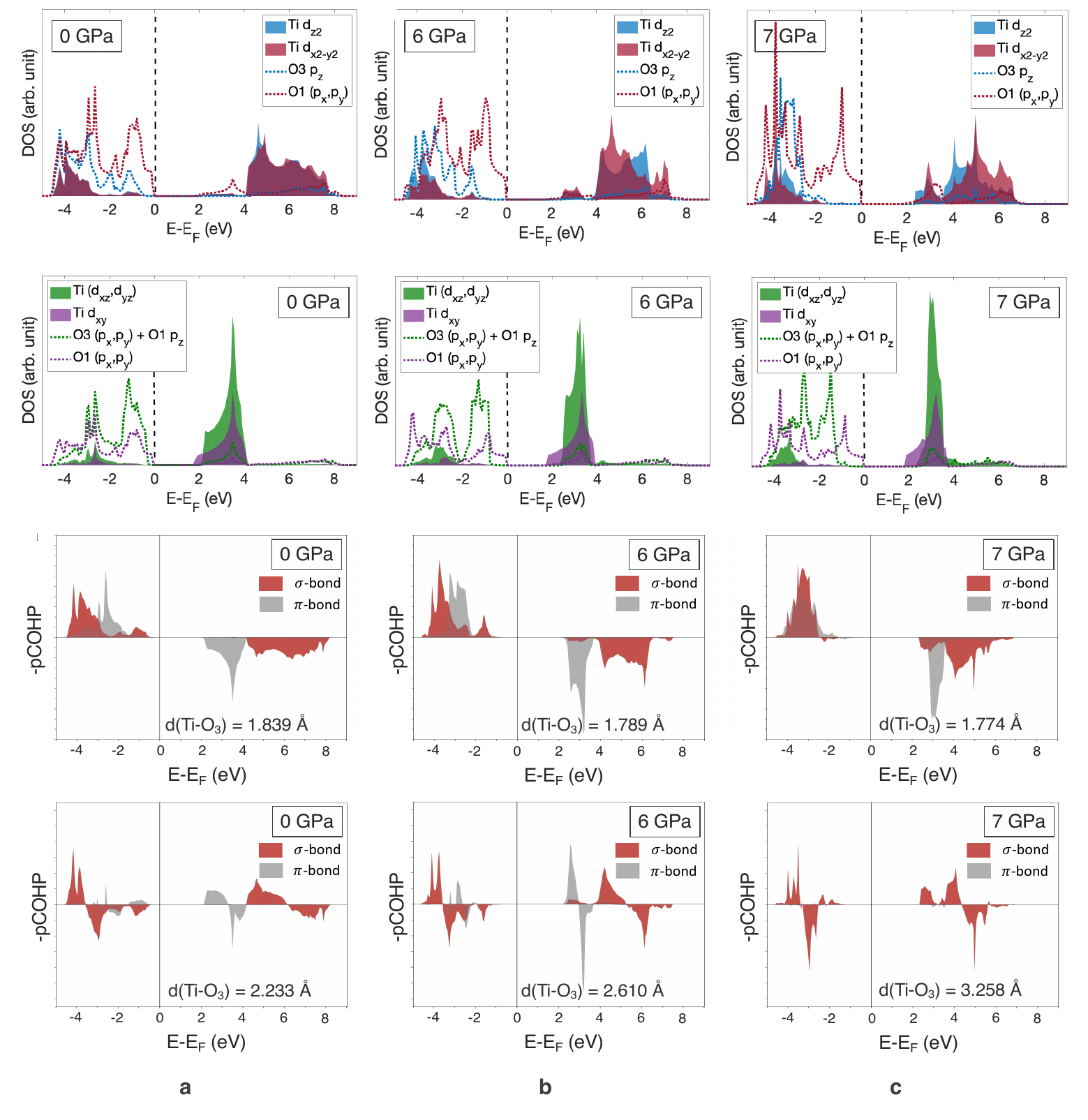}\caption{Partial density of states and the projected crystal orbital hamiltonian
populations (pCOHP). The partial density of states involved in the
$\sigma$-bond (top row) and in the $\pi$-bond (second row) and the
pCOHP evaluated along the short (third row) and long (last row) $\ce{Ti-O }$
bond of (a) $0$ GPa, (b) $6$ GPa and (c) $7$ GPa. For the pCOHP,
the red filling represents the $\sigma$-bond contribution ($\ce{Ti }$
$d_{z^{2}}$ with $\ce{O_{3} }$ $p_{z}$ was taken), and the gray
filling represents the $\pi$-bond contribution ($\ce{Ti }$ $d_{yz}$
with $\ce{O_{3} }$ $p_{y}$ was taken).}
\label{fig:DOS}
\end{figure*}

Hence, the charge density was studied within a plane wave formalism
as well as within the projected crystal orbital hamiltonian population
(pCOHP) theory to obtain a more detailed picture of the chemical changes
that occurs in the isosymmetric phase transition. For the equilibrium
tetragonal phase, it can be seen in the DOS (see Figure \ref{fig:DOS}a)
that there is a crystal field splitting of the $e_{g}$ and $t_{2g}$
orbitals that correspondingly form the $\sigma$-bond and $\pi$-bond,
with the bonded orbitals between $-5$ eV and $-2$ eV. The valence
band maximum (VBM) largely consists of unbonded oxygen $p$-orbitals
(forming the $t_{1u}$ and $t_{2u}$ orbital levels), while the antibonded
states of the $\pi$-bond ($t_{2g}^{*}$) form the conduction band
minimum (CBM). Upon transitioning, there are several significant changes
in the DOS that can be identified.

\ 

Firstly, the $\sigma$-bonding $d$-orbitals remain nearly unchanged
throughout the whole pressure range, indicating their minor role in
the ST phase transition. On the other hand, the width of the $(d_{xz},d_{yz})$
orbital DOS weight decreases, which implies the localization of the
electrons within these $\pi$-bonded $d$-orbitals. Secondly, these
latter orbitals are also shifted towards lower energies within the
valence band, such that they coincide within the energy range of the
$e_{g}$ orbital levels. Thirdly, the VBM consisting of non-bonding
oxygen $p$-orbitals also changes drastically with a vanishing weight
of the $\ce{O }_{3}$ atom, indicating that this atom is fully involved
in the bond with titanium upon the ST transition.

\ 

However, a complete chemical interpretation of these changes in the
electronic DOS remains a delicate task, therefore, the pCOHP was evaluated.
This is the DOS weighted by its Hamiltonian matrix element, indicative
of the bond strength. These calculations were performed on a $1\times1\times2$
supercell in order to properly compare the pCOHP of the short and
long $\ce{Ti-O }$ bond along the polar axis. In this way the bonding
(negative pCOHP) and antibonding (positive pCOHP) states that are
formed in the tetragonal distortion can be clearly seen in the pCOHP
(bottom two rows in Figure \ref{fig:DOS}a). For the short $\ce{Ti-O }$
bond around the transition, it is again observed that the $\pi$-bond
weight (both for bonding as antibonding states) has a reduced width,
while the bonded states shift to lower energies coinciding with the
$\sigma$-bonded weight. The first feature indicates that the electron
are localized into well-defined hybrid orbital states, which is in
agreement with the observations in section \ref{subsec:Dielectric-Properties}.
The latter feature involves the combination of two effects, namely
an enhanced crystal field splitting as well as an increased bond strength
(orbital overlap).

The pCOHP of the long $\ce{Ti-O }$ bond looks very different with
nearly equal amount of weight of bonding and antibonding states below
the Fermi level, which implies a bond order of zero. Most interestingly,
one observes a vanishing of the $\pi$-bond pCOHP weight upon the
transition to the ST phase at $7$ GPa. This implies that there is
no orbital overlap of the $(d_{xz},d_{yz})$ along the long bond axis,
while the $\sigma$-bond weight remains. Additionally, no change in
the in-plane components of the pCOHP could be inferred (see Supplemental
Information). Hence, it can be concluded that the ST first-order phase
transition has been mainly driven by a broken $\pi$-bond along the
long $\ce{Ti-O }$ bond.

\begin{figure*}
\includegraphics[scale=0.65]{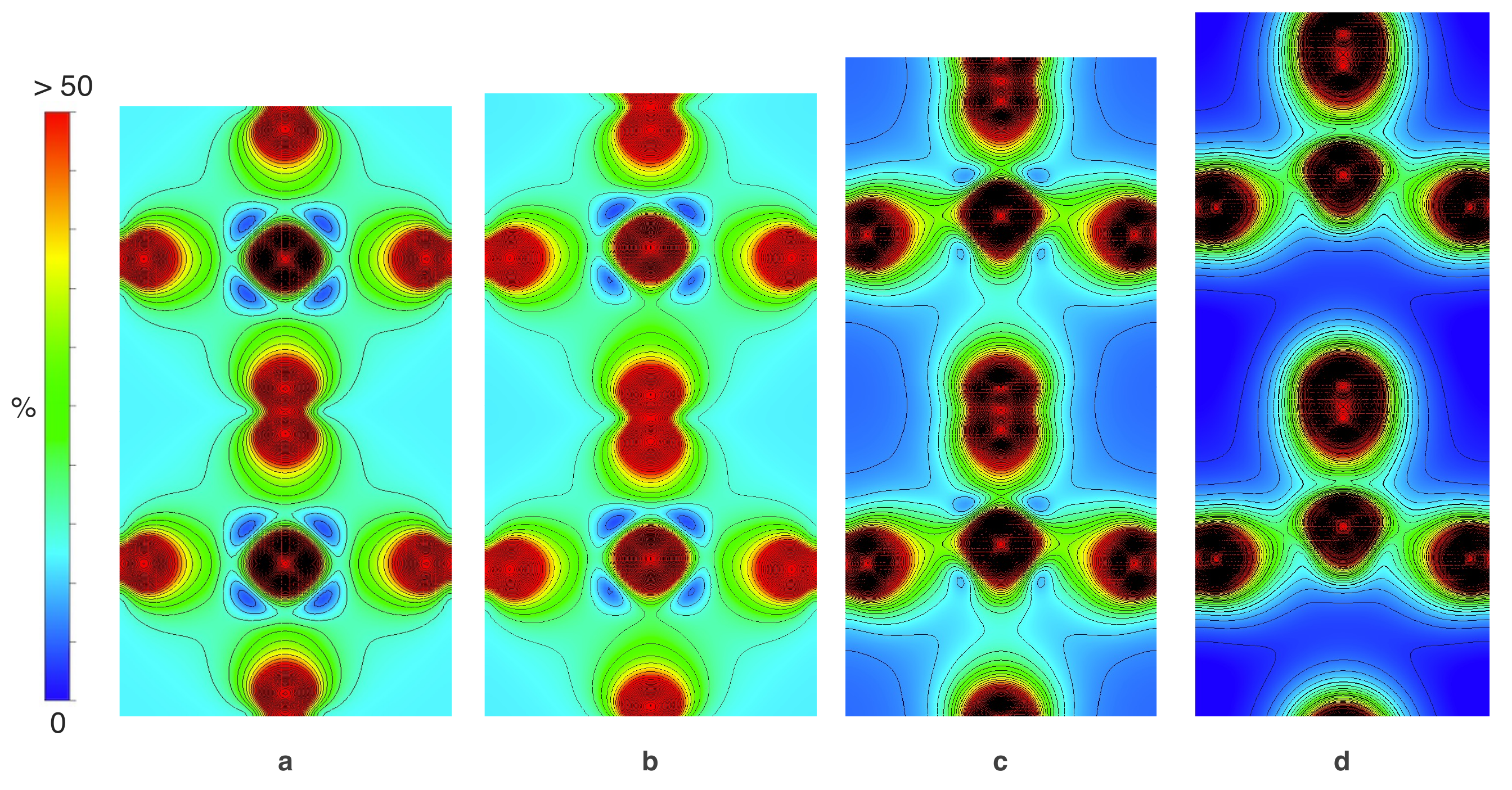}\caption{Projected valence band charge density. The valence band charge density
taken along the $(200)$ plane in the unit cell of the (a) cubic,
(b) tetragonal, (c) tetragonal at $6$ GPa, and (d) at $7$ GPa.}
\label{fig: VCD}
\end{figure*}

These observations are confirmed visually by the calculation of the
valence charge density (VCD) taken along the $(200)$ plane, as shown
in Figure \ref{fig: VCD}. First, a comparison of the VCD of the cubic
and tetragonal phase shows the subtle change in the charge density
(See Fig. \ref{fig: VCD}a-b), which can mosty easily be seen by an
increased density of the isosurface lines indicating the enhanced
covalency upon FE distortion. Upon increasing the negative pressure,
the VCD decreases and increases along the long and short $\ce{Ti-O }$
bond, respectively. In particular, the $\ce{O }_{3}$ anion gains
in density perpendicular to the short $\ce{Ti-O }$ bond, indicating
enhancement of the $\pi$-bond while the empty $(d_{xz},d_{yz})$
lobes of $\ce{Ti }$ are fully depleted, implying a strong hybridization
with the $\ce{O }$ $2p$ orbitals.

These findings offers guidance in the understanding of first-order
isosymmetric phase transitions \citep{CeriumVolumeCollapse,Isosymmetric1,Isosymmetric2,Pia_pinningMott},
which turn out to be often characterized by an anisotropic structural
change (ascribed to the high-directionality of the covalent bond).
These phase transitions appear in materials where a small anistropy
already exists - often a transition metal off-centring from a high-symmetric
position. They form the seed of isosymmetric phase transitions, as
they inhibit the covalent bond whose anisotropy can be abruptly enhanced
by an external trigger giving rise to a structural first-order transition.
Additionally, this sudden enhancement of the existing bond states
can also be driven by electron correlations, such as in the case of
the Mott metal-insulator transition in $\ce{V_{2}O_{3} }$ where inducing
a change in the orbital overlap of the $d_{z^{2}}$ triggers an electronic
repulsive force between the neighboring vanadium cations \citep{Pia_pinningMott}.

\section{\label{sec:Conclusion} Conclusions}

Our results proof the existence of a ST phase of $\ce{BaTiO_{3} }$
upon negative pressure of $7$ GPa, with an enhanced polarization
of $\sim95\:\mu$C/cm$^{2}$. It was shown that a biaxial strain can
not fully induce the ST phase, indicating the opportunities that appear
by controlling all three lattice parameters \citep{3DstrainGe}. The
microscopic origin of this first-order structural phase transition
involves a charge re-ordering in which the $\pi$-bond along the long
$\ce{Ti-O }$ bond completely vanishes while strong $\pi$-bonds
are formed in a five-fold coordination, preserving the $\ce{Ti }^{4+}$
oxidation state. It can be roughly said that the negative pressure
opens a new avenue for the electrons to become localized in well-defined
covalent bonds along an exceptionally short $\ce{Ti-O }$ bond, instead
of localizing these electrons by additional covalent bonds formed
with two ($Amm2$) or three ($R3m$) oxygen atoms. These findings
shed light on how isosymmetric phase transitions are driven by subtle
changes in their chemical bonding, requiring the need for a local
auxiliary atomic orbital description.

\section{\label{sec:Methods}Computational Methods}

All density functional theory (DFT) calculations were performed by
the Vienna \emph{ab initio} simulation package (VASP) \citep{VASP}.
The interactions between electrons and ions were described by the
projector augmented wave (PAW) potentials \citep{PAWmethod}, with
the electronic wave functions expanded with a large cutoff energy
of $900$ eV. The PBEsol exchange-correlation functional \citep{PBEsol} was
used for all calculations. For the structural relaxation, a force
convergence criterion of $0.005$ eV/$\AA$ was used with the
Brillouin zone (BZ) sampled by a $12\times12\times12$ $\Gamma$-centered
$k$-scheme. Phonon dispersions and dielectric properties were calculated
self-consistently on the basis of density functional perturbation
theory (DFPT) and with the use of the PHONOPY package \citep{phonopy}.
For the electronic and dielectric properties a denser $24\times24\times24$
mesh was used with energy convergence criterion of $10^{-6}$ eV.
The Berry phase method as implemented in VASP was adopted for the
calculation of the polarization \citep{Polarization1,Polarization2}.
For the crystal orbital hamiltonian population (pCOHP), the LOBSTER
package \citep{Lobster,pCOHP_lobster} was used with basis sets $\ce{Ba }$
$(5s,5p,5d,6s)$, $\ce{Ti }$ $(3s,3p,3d,4s)$ and $\ce{O }$ $(2s,2p)$
with a minimal charge spilling. 

\section*{Acknowledgements}

Part of this work was financially supported by the KU Leuven Research
Funds, Project No. C14/21/083, iBOF/21/084, No. KAC24/18/056 and No.
C14/17/080 as well as the Research Funds of the INTERREG-E-TEST Project
(EMR113) and INTERREG-VL-NL-ETPATHFINDER Project (0559). Part of the
computational resources and services used in this work were provided
by the VSC (Flemish Supercomputer Center) funded by the Research Foundation
Flanders (FWO) and the Flemish government.

\section*{Author contributions}

These results were obtained within the framework of the doctoral training
of S.M. supervised by J.-P.L., M.H. and J.W.S. S.M. performed all
calculations and wrote the manuscript with contributions and comments
from all authors.

\section*{Competing interests}

The authors declare no competing interests.

\pagebreak
\widetext
\begin{center}
\textbf{\large Supplemental Material}
\end{center}

\subsection{Structure of $\ce{BaTiO_{3} }$ Phases Under Negative Pressure}

\begin{figure}[h]
\includegraphics[scale=0.60]{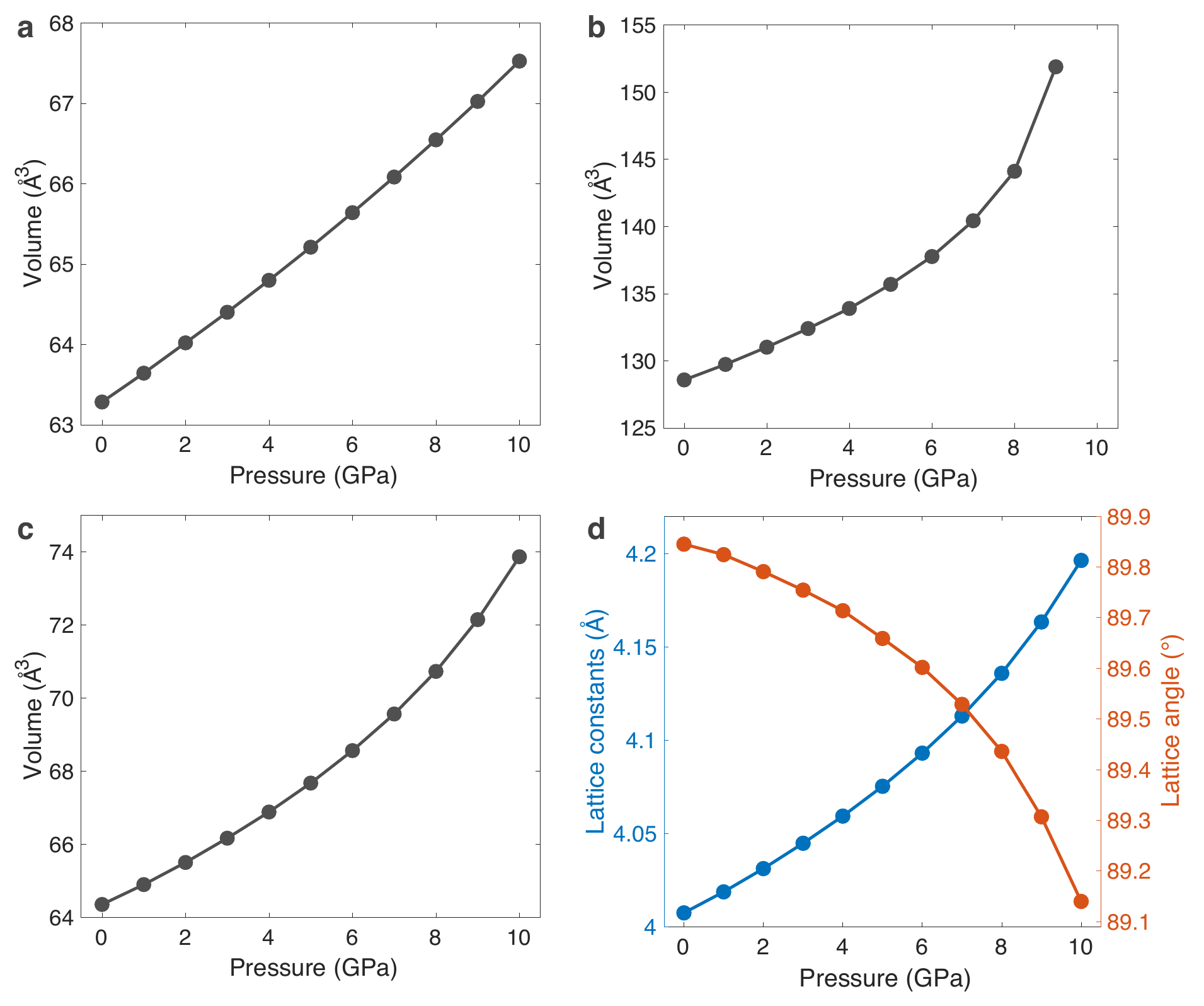}\caption{Structure of $\ce{BaTiO_{3} }$ phases under negative pressure. The
volume under negative pressure of the (a) cubic (b) orthorombic and
(c) rhombohedral under negative pressure. (d) The structural parameters
of the rhombohedral phase.}

\end{figure}

\subsection{Enthalpy Comparison }

\begin{figure}[h]
\includegraphics[scale=0.33]{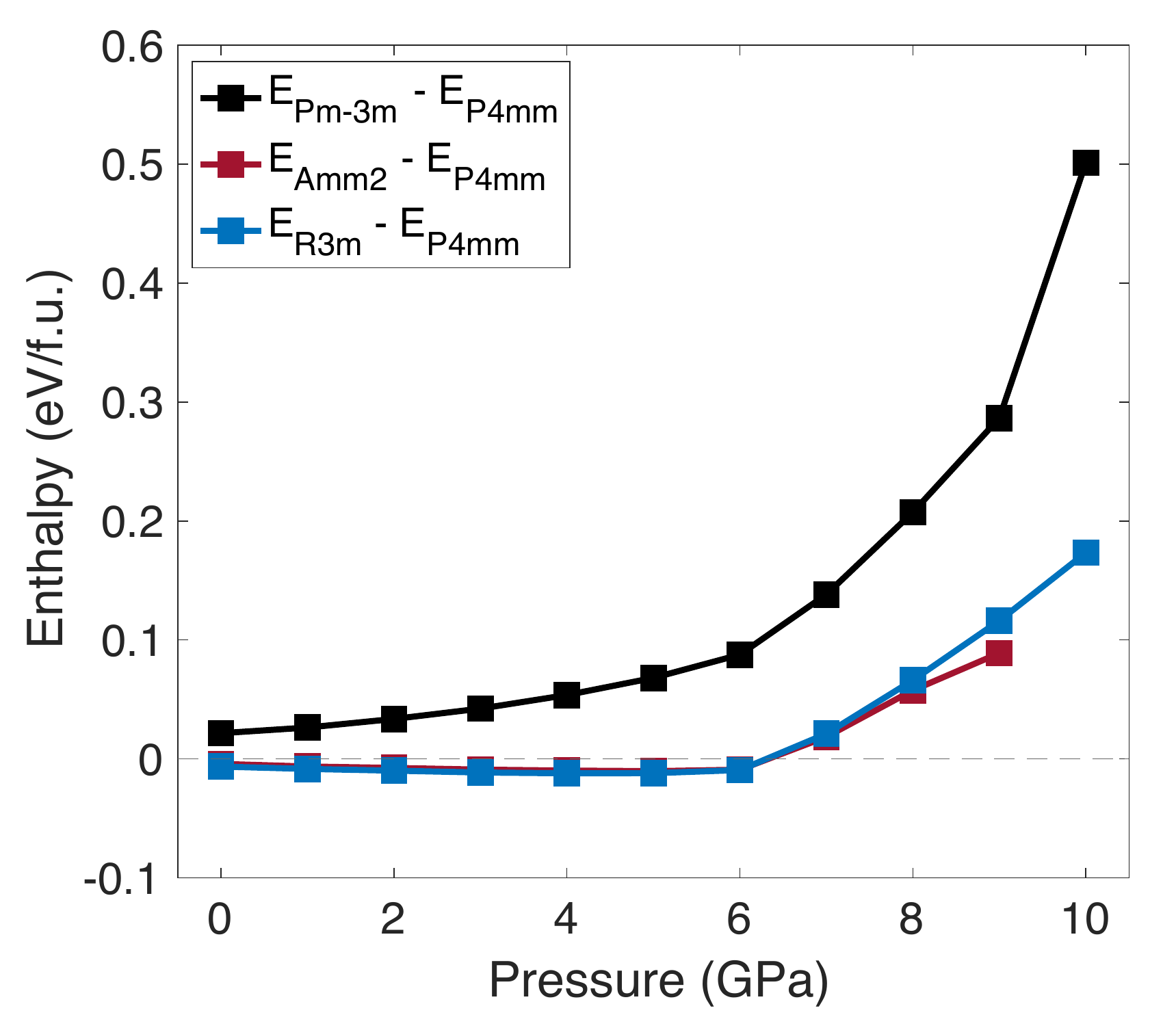}\caption{Enthalpy comparison of the cubic ($Pm\bar{3}m$), tetragonal ($P4mm$),
orthorombic ($Amm2$) and rhombohedral ($R3m$).}
\end{figure}

\subsection{Atomic charge analysis}

\begin{table}[h]
\begin{tabular}{|c||c|c|c||c|c|c|}
\hline 
 & $0$ GPa & $6$ GPa & $7$ GPa & $0$ GPa & $6$ GPa & $7$ GPa\tabularnewline
\hline 
\hline 
$\ce{Ba }$  & $1.568$ & $1.569$ & $1.574$ & $2.11$ & $1.96$ & $1.83$\tabularnewline
\hline 
$\ce{Ti }$  & $2.129$ & $2.109$ & $2.119$ & $0.82$ & $1.00$ & $1.19$\tabularnewline
\hline 
$\ce{O_{1} }$  & $-1.240$ & $-1.249$ & $-1.256$ & $-0.99$ & $-0.80$ & $-1.03$\tabularnewline
\hline 
$\ce{O_{3} }$  & $-1.218$ & $-1.180$ & $-1.187$ & $-0.95$ & $-0.93$ & $-0.95$\tabularnewline
\hline 
\end{tabular}\caption{The Bader atomic charges (left) and M\"ulliken charges (right).}

\end{table}

\subsection{Electron localization function}

\begin{figure}[h]
\includegraphics[scale=0.40]{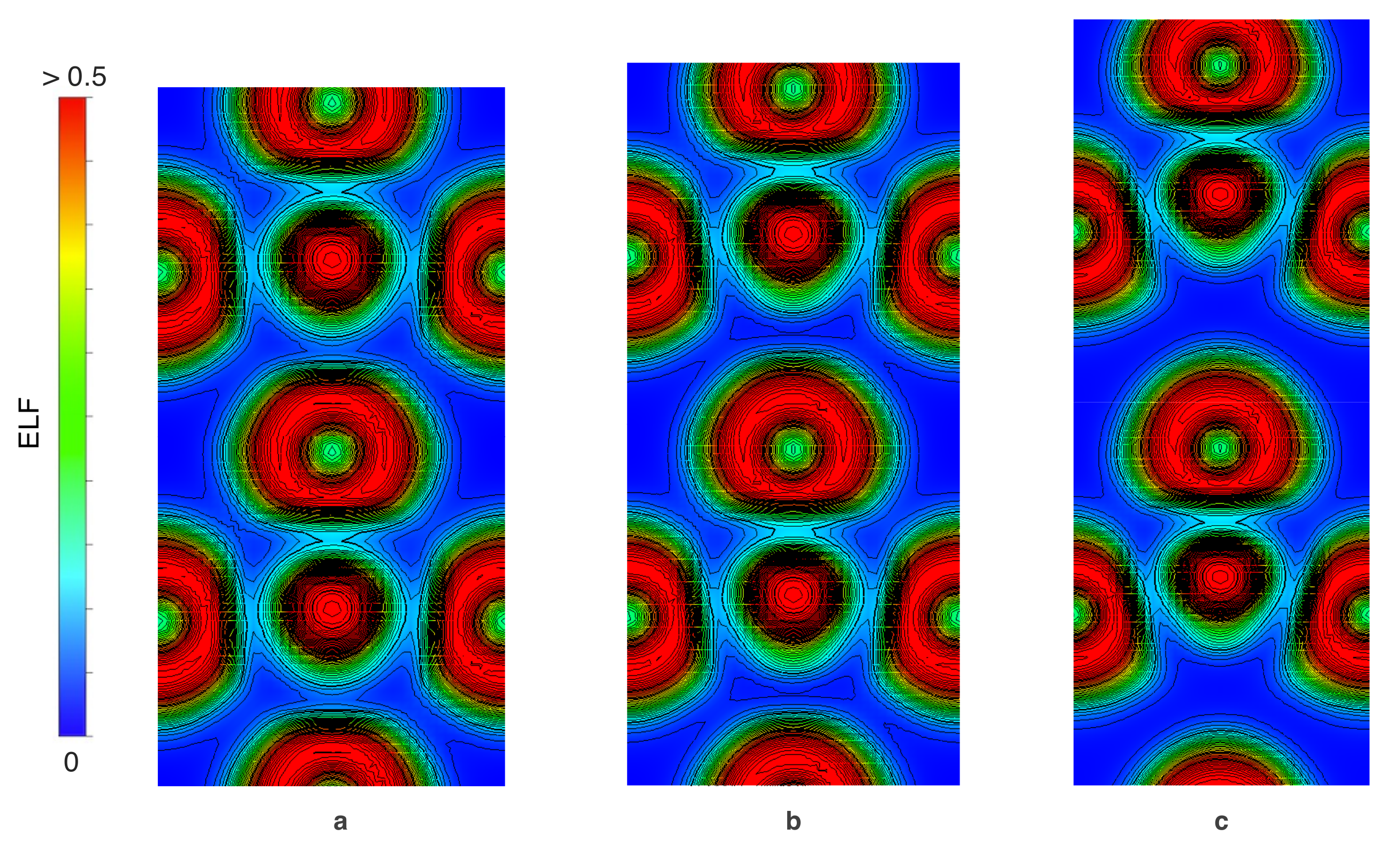}\caption{Electron localization function as a function of pressure. ELF taken
along the $(200)$ plane for (a) $0$ GPa, (b) $6$ GPa and (c) $7$
GPa.}

\end{figure}

\subsection{In-plane pCOHP}

\begin{figure}[h]
\includegraphics[scale=0.58]{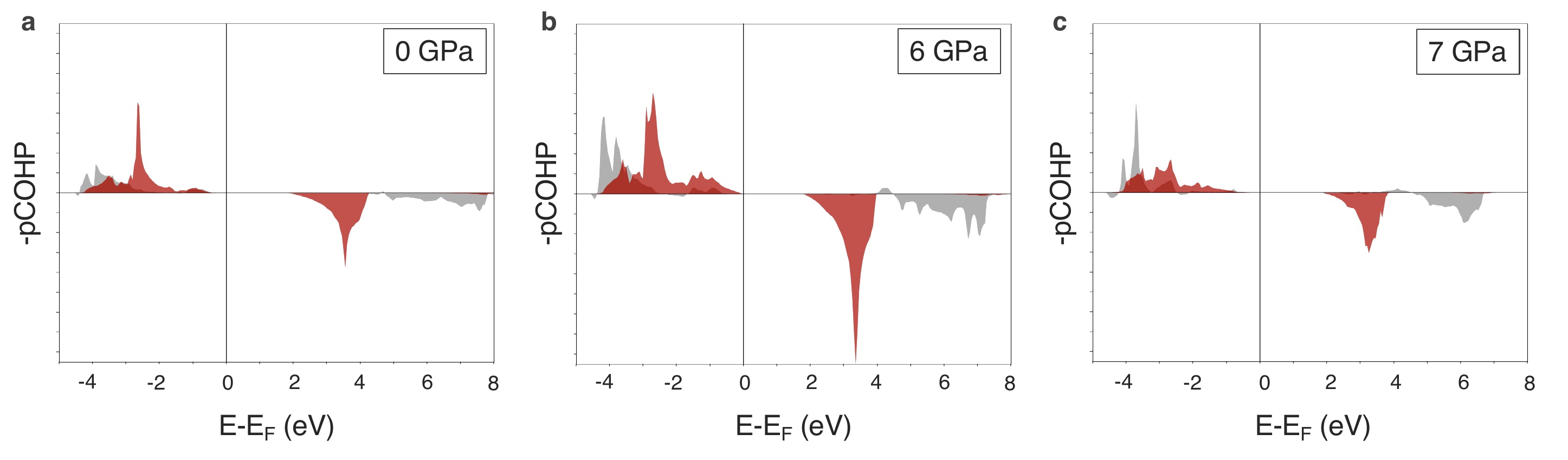}\caption{The projected crystal orbital hamiltonian population (pCOHP) of the
in-plane bonds. Red and grey shaded area representing the $\sigma$-bond
($\ce{Ti }$ $d_{x^{2}-y^{2}}$ with $O_{1}$ $p_{x}$ was taken)
and the $\pi$-bond ($\ce{Ti }$ $d_{xy}$ with $O_{1}$ $p_{y}$
was taken), respectively. }

\end{figure}

\bibliographystyle{unsrt}
\bibliography{referencesBTO}

\end{document}